# Influence of global correlations on central limit theorems and entropic extensivity

John A. Marsh[1], Miguel A. Fuentes[2,3],
Luis G. Moyano[4] and Constantino Tsallis[2,4]

[1]SI International, 7900 Turin Road, Rome, NY 13440, USA
[2]Santa Fe Institute, 1399 Hyde Park Road, Santa Fe, NM 87501, USA
[3]Consejo Nacional de Investigaciones Científicas y Técnicas, Centro Atómico Bariloche and Instituto Balseiro, 8400 Bariloche, Rio Negro, Argentina
[4]Centro Brasileiro de Pesquisas Físicas, Rua Xavier Sigaud 150, 22290-180 Rio de Janeiro, Brazil

**Abstract**
We consider probabilistic models of $N$ identical distinguishable, binary random variables. If these variables are strictly or asymptotically independent, then, for $N \to \infty$, (i) the attractor in distribution space is, according to the standard central limit theorem, a Gaussian, and (ii) the Boltzmann-Gibbs-Shannon entropy $S_{BGS} \equiv -\sum_{i=1}^{W} p_i \ln p_i$ (where $W = 2^N$) is extensive, meaning that $S_{BGS}(N) \sim N$. If these variables have any nonvanishing *global* (i.e., not asymptotically independent) correlations, then the attractor deviates from the Gaussian. The entropy appears to be more robust, in the sense that, in some cases, $S_{BGS}$ remains extensive even in the presence of strong global correlations. In other cases, however, even weak global correlations make the entropy deviate from the normal behavior. More precisely, in such cases the entropic form $S_q \equiv \frac{1}{q-1}\left(1 - \sum_{i=1}^{W} p_i^q\right)$ (with $S_1 \equiv S_{BGS}$) can become extensive for some value of $q \neq 1$. This scenario is illustrated with several new as well as previously described models. The discussion illuminates recent progress into *q-describable* nonextensive probabilistic systems, and the conjectured *q*-Central Limit Theorem (*q*-CLT) which posses a *q*-Gaussian attractor.

## 1. Introduction

The foundation of statistical mechanics is a probabilistic description of macroscopic systems reflecting microscopic dynamical behavior. If the assumption of ergodicity holds in configurational phase space, then we have standard Boltzmann-Gibbs statistical mechanics. Chaos theory provides the basic tools and simple models to study situations in which ergodicity may be violated, showing fractal, hierarchical, or other *incomplete* occupation of phase space. A generalization of the Boltzmann-Gibbs-Shannon entropy $S_{BGS}$ may provide a fundamental description characterized by a single real parameter $q$. The entropy $S_q$ is defined, for a system with $W$ microstates with occupation probabilities $p_i$, as follows:

$$S_q \equiv \frac{1}{q-1}\left(1-\sum_{i=1}^{W} p_i^q\right) \quad \text{(with } S_1 = S_{BGS} \equiv -\sum_{i=1}^{W} p_i \ln p_i \text{)} \qquad (1)$$

For composition of *independent* subsystems, where the joint probabilities satisfy a product law, $S_q$ is extensive for $q = 1$, and nonextensive for $q \neq 1$. In many respects, the appearance of nonextensivity provides an essential description of natural, man-made, and even social phenomena. For example, in physical systems, long-range interactions (and, hence, global correlations) among system components, or long-range memory, imply that the composition of systems cannot proceed using the assumption of probabilistic independence.

The statistical mechanics associated with $S_q$, which parallels that associated with Boltzmann-Gibbs entropy, has been developed to a considerable degree [1], and gives rise to the so-called *nonextensive statistical mechanics*, with stationary state distribution function showing power-law ("heavy-tail") decay. Strong connections have also been made by Robledo and collaborators [2] (see also [3]) between the $q$-statistical concepts and nonlinear dynamical systems such as unimodal one-dimensional maps at the edge of chaos and elsewhere.

Recent developments [4,5] have provided a number of simple examples which illustrate the interconnected nature of these concepts and the applicability of this formalism to natural systems, in which strongly correlated behaviors and power-law distributions are ubiquitous, yet largely unexplained.

The aim of this communication is to explore globally correlated systems composed of $N$ distinguishable and identical random binary variables. Two aspects of these systems are of particular interest: first, the question of extensivity of the entropy $S_q$, and second, the nature of the attractor in the space of probability distributions. Each of these aspects may show anomalous behavior due to global correlations, in contrast to the more normal behavior associated with strictly or nearly independent system composition. The entropy $S_q$ which is extensive in the case of independence is that with $q = 1$, whereas the anomalous case shows extensivity for $q \neq 1$. Correspondingly, for independent systems, the attractor in probability space is a Gaussian, as studied by A. de Moivre (1733), P.S. de Laplace (1774), R. Adrain (1808) and C.F. Gauss (1809), when the variance of the single distribution is finite. If this variance diverges instead, the attractor is a Levy distribution, as studied by P. Levy and B.V. Gnedenko in the 1930's. In the anomalous case (i.e., when global correlations are present), a variety of attractors have been conjectured or discussed [6], most notably the $q$-Gaussian distribution [5,7]. This distribution optimizes the entropy $S_q$, and so the two anomalous features induced by global correlations are intimately related. Several examples of globally correlated systems will be studied here in this context, and the unified picture that emerges will be discussed in the light of recent progress [5,7] at a $q$-generalized central limit theorem ($q$-CLT).

This new formalism has also been shown [8] to yield a fundamental explanation for the emergence of networks in nature: network nodes are interpreted as special points (or regions) in a more-or-less continuous phase space which show non-vanishing occupancy.

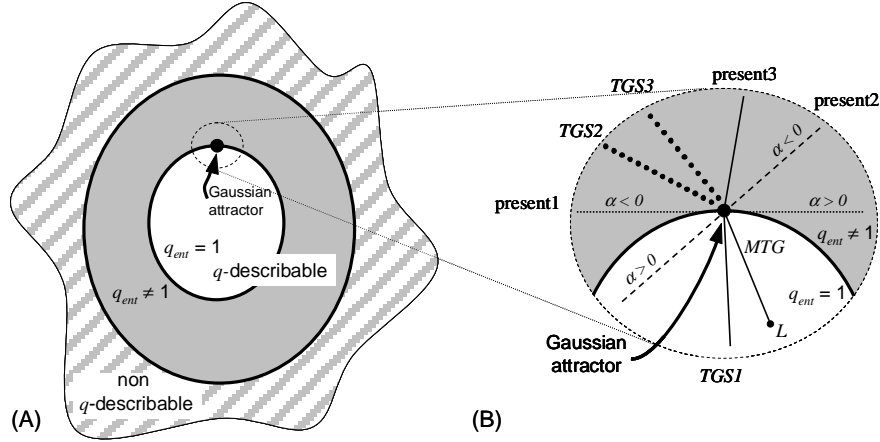

Fig. 1: (A) Schematic representation of the classes of probabilistic systems discussed. The class of $q$-describable systems is defined as those for which the entropy $S_q$ is extensive for some $q = q_{ent}$. The $q$-describable systems are further divided into those for which the Boltzmann-Gibbs-Shannon entropy $S_{BGS} = S_1$ is extensive, and those for which $S_q$ is extensive for $q_{ent} \neq 1$. The Gaussian attractor occurs for the special case of independent system composition, and is represented here by a single point contained within the space of $q_{ent} = 1$ $q$-describable systems. (B) Detail showing the various models discussed in this work, appearing as lines parameterized by some variable in the model. The positioning of the lines indicates whether $q_{ent}$ equals unity or differs from unity. The various models are labeled as discussed in the text: the MTG endpoint $L$ represents the Leibnitz triangle.

The simplest setting in which to study correlated behavior of a large number of subsystems is discrete $k$-state systems under composition. (We note that analogous constructions can be made for continuous systems [4]). As $N$ $k$-state systems are composed, the number of possible states is given by $W = k^N$. Let us denote by $W^{eff}$ the number of states whose probability is (sensibly) different from zero (*eff* stands for *effective*). If the subsystems are independent, then generically $W^{eff} = W = k^N$. If, furthermore, each of these possibilities has probability on the order of $1/k^N$, the BGS entropy is given by $S_{BGS} \sim \ln W \sim N \ln k$, and thus shows extensive behavior (i.e., $S_{BGS} \sim N$ as $N \to \infty$, or, equivalently, $0 < \lim_{N \to \infty} \frac{S_q}{N} < \infty$) only for $q_{ent} = 1$. Thus, the composition of (strictly or nearly) uncorrelated systems is described by the usual entropy. However, if strong correlations exist among subsystems, then, in general, many microstates of the system might be forbidden. For a large class of interesting systems, the effective number of states $W^{eff} \sim N^\rho \ll W = k^N$ ($\rho > 0$), in which case

$$S_q \sim \ln_q W^{eff} \equiv \frac{(W^{eff})^{1-q}-1}{1-q} \sim N^{\rho(1-q)}. \qquad (2)$$

In this case $S_q$ is extensive for and only for

$$q_{ent} = 1 - \frac{1}{\rho}. \qquad (3)$$

This simple argument links a (appreciably) reduced occupancy of phase space, corresponding to a breaking of ergodicity, with the entropy $S_q$ and the associated nonextensive theory. It becomes clear now that the expression "nonextensive entropy" sometimes used for $S_q$, is appropriate only for "normal" systems (i.e., those constituted by subsystems which are mutually independent), not in general. For such systems, the multiplicity of microstates is exponential in the number $N$ of composed subsystems. Focus has now turned instead to the so-called "$q$-describable" systems [4, 9] in which the entropy $S_q$ shows extensive behavior for some $q_{ent}$ equal to or different from unity. Fig. 1a shows a schematic representation of these possibilities, with $q$-describable systems forming a subset of all possible systems, and $q$-describable systems with $q_{ent} = 1$ forming a further subset of the $q$-describable systems. The "normal", or independent, case is shown as a point located at the boundary of the $q_{ent} = 1$ and $q_{ent} \neq 1$ regions of the $q$-describable systems. We will return to Fig. 1 in the next section, when specific models are discussed.

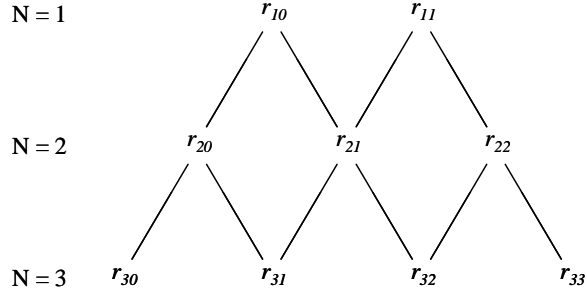

Fig. 2: Pascal-Leibnitz triangle configuration of reduced probabilities $r_{N,n}$ illustrated for number of composed systems $N$ = 1, 2, 3.

Our interest here lies not only in simple examples of $q$-describable systems, but also some indications as to whether $q$-statistics (with the associated power-law generalizations of the natural logarithm and exponential functions) constitutes a fundamental explanation of (asymptotic) scaling and power-law behavior observed so often in nature. Optimizing $S_q$ in the new formalism, with appropriate constraints, yields the so-called $q$-Gaussian, a generalized form of the Gaussian distribution, given by $f(x) \propto e_q^{-\beta x^2}$, where $\beta$ characterizes the width, and the $q$-exponential function $e_q^x \equiv [1+(1-q)x]^{1/(1-q)}$ (so defined where $1-(1-q)x > 0$, and 0 elsewhere) appears, and reduces, as expected, to the usual exponential function in the limit $q = 1$. This $q$-Gaussian distribution has power law tails for $1 < q < 3$, has compact support for $q < 1$, and recovers the usual Gaussian distribution for $q = 1$. From this perspective, we expect the $q$-Gaussian to appear often in nature, associated with strongly correlated complex systems. Indeed, a considerable amount of evidence exists today for the ubiquitous nature of power-law behavior in both natural and man-made systems, and the possibility that the $q$-statistical formalism is in some sense fundamental in nature is thus very appealing.

Important evidence in support of such a possibility has recently appeared in the form of numerical indications [5, 7], and a formal development of a $q$-qeneralized Central Limit Theorem ($q$-CLT) is in progress.

A natural starting point is the case of identical and distinguishable binary subsystems, but which are not necessarily independent. We let $p$ denote the probability that the binary system is in state 0, out of a state space $\{0,1\}$. When $N$ such systems are composed, the full state space may be considered to consist of all $2^N$ possible binary strings of length $N$. Because the systems are identical, these $2^N$ probabilities can be reduced to just $N$ values $r_{N,n}$, each with multiplicity given by the binomial coefficient $N!/[(N-n)!n!]$, where $n = 0$, 1, ..., $N$. These *reduced probabilities* $r_{N,n}$ thus satisfy the normalization condition

$$\sum_{n=0}^{N} \frac{N!}{(N-n)!\,n!} r_{N,n} = 1 \tag{4}$$

and $r_{N,n}$ can be thought of as counting the number of subsystems in the state 1 (and thus $r_{1,0} \equiv p$ and $r_{1,1} \equiv 1-p$). Any specific prescription for successively adding binary systems to the $N$-system yields a sequence of systems, whose reduced probabilities $r_{N,n}$ can be arranged into a triangle, as shown in Fig. 2. The corresponding arrangement of the binomial coefficients is the familiar Pascal triangle. Since each of the $N$ single random variables can take values 0 or 1, the index $n$ corresponds to the value of the random variable defined by the *sum of all $N$ binary random variables*. Therefore, the attractors that emerge after centering and rescaling *n, precisely correspond to the natural abscissa associated with all central limit theorems.* The abscissa $n$ measured from its "center" (value of $n$, for given $N$, for which $p_{N,n}$ is maximal) scales, in all the models that we discuss in the present paper, as $N^\gamma$ with $\gamma \geq 0$. Therefore, $\gamma = \frac{1}{2}$ corresponds to *normal diffusion*, $\gamma > \frac{1}{2}$ corresponds to *superdiffusion*, and $0 \leq \gamma < \frac{1}{2}$ corresponds to *subdiffusion* ($\gamma = 0$ means in fact *localization*).

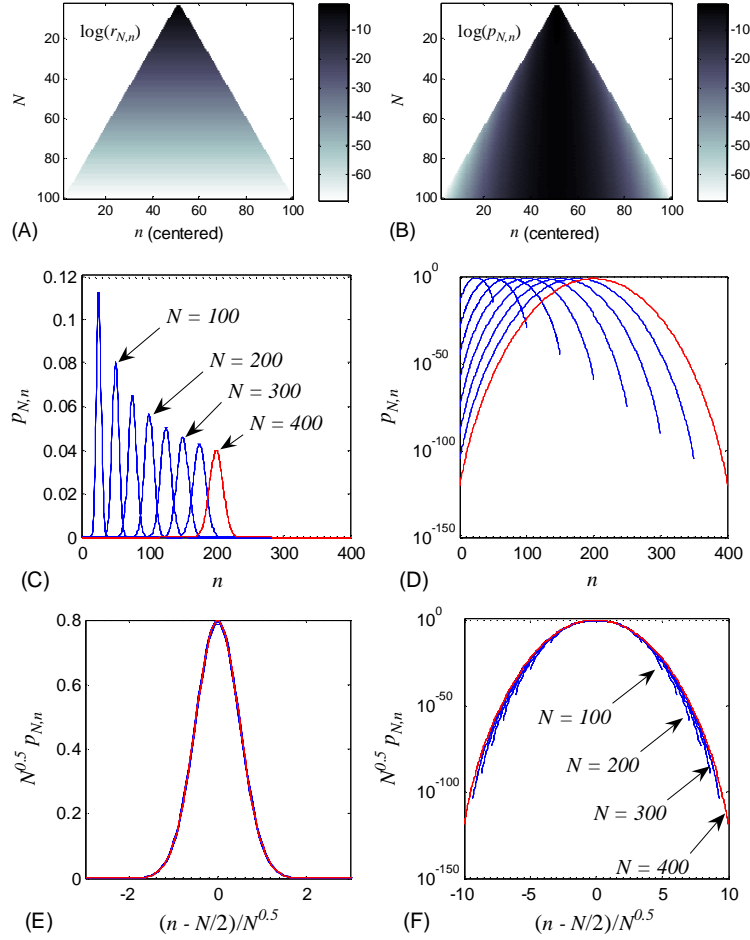

Fig. 3: Behavior of independent binary system composition for the case $p = 0.5$. Each model discussed in this paper has this case as a limiting case. Parts (A)-(B) show reduced probabilities $\log(r_{N,n})$ and probabilities $\log(p_{N,n})$ up to $N = 100$ as grayscale images. Parts (C)-(D) show uncentered and unscaled probabilities for $N = 50, 100, \ldots, 400$, in both linear and log units. Parts (E)-(F) show centered and scaled probabilities, illustrating the Gaussian attractor in probability space.

A scale-invariant construction, yielding systems suitable for analysis in the thermodynamic limit $N \to \infty$, can be imposed by requiring the marginal probabilities of the system at level $N$ in the construction to reproduce the joint probabilities at level $N - 1$. It can be verified that this condition is equivalent to imposing the *Leibnitz rule* for the reduced probabilities: $r_{N,n} + r_{N,n+1} = r_{N-1,n}$.

When satisfied, the Leibnitz rule provides a prescription for determining all reduced probabilities given just one value at each level $N$ in the construction, say $p_{N,0}$. We note that imposition of the Leibnitz rule ensures conservation of probabilities as in Eq. 4, however not all choices of $p_{N,0}$ will yield *admissible* probability sets in which $0 \leq p_i \leq 1, \forall i$.

In what follows, we first summarize several examples of correlated systems that have appeared recently in the literature, and then present a few new models that show non-trivial entropy growth. The discussion that follows focuses on the broad classification of probabilistic systems afforded by these considerations, highlighting some questions that remain unanswered.

## 2. Models

We shall present now several models which illustrate various cases of entropy growth, and attractor in phase space. Each model is shown schematically in Fig. 1B as either a continuous or dotted line, representing continuously or discretely parameterized models, respectively. All the models we consider are $q$-describable, and the position of the model in Fig. 1A indicates whether $q_{ent} = 1$ or $q_{ent} \neq 1$. The models each reduce to the independent ("normal") case at some value (or limiting value) of the parameter(s), and therefore each model either terminates at or passes through the point representing the independent case (labeled "Gaussian attractor") in the figure.

For each model, we will first provide a definition by specifying the reduced probability set $r_{N,n}$, then characterize the model by considering the entropy growth and the large $N$ limit of the attractor. In terms of the reduced probabilities $r_{N,n}$, the entropy as a function of $N$ is given by

$$S_q(N) \equiv \frac{1}{q-1}\left(1 - \sum_{i=1}^{2^N} p_i^q\right) = \frac{1}{q-1}\left(1 - \sum_{n=0}^{N} \frac{N!}{(N-n)!n!}(r_{N,n})^q\right), \quad (5)$$

which makes explicit the degeneracy of the reduced probabilities.

The case of independent binary system composition will serve as a model for the analysis of the other models presented here. Fig. 3 illustrates reduced probabilities $r_{N,n}$ and probabilities $p_{N,n}$ for the independent case where $p = 0.5$, and for various $N$, where

$$p_{N,n} \equiv r_{N,n} \frac{N!}{(N-n)!\,n!}. \quad (6)$$

The grayscale plots (A) and (B) show $r_{N,n}$ and $p_{N,n}$, respectively, centered so the values appear in a triangle. The plots (C) and (D) show the unscaled and uncentered values $p_{N,n}$ in both linear and log units. Finally, plots (E) and (F) show $p_{N,n}$ scaled and centered such that the Gaussian attractor is clearly observed. Indeed, as seen in the axis labels, the scaling makes use of the fact that the centers of the unscaled distributions occur at values $N/2$, and the widths scale as $\sqrt{N}$. In general, for the models presented here, the attractor in phase space is scaled by first centering, and then applying the rule

$$p_{N,n} \to p_{N,n} N^{\gamma}$$
$$n \to \frac{(n - n_{center})}{N^{\gamma}}, \quad (7)$$

which generalizes the scaling used for the independent (Gaussian) case, where the exponent was $\gamma = 0.5$. The exponent $\gamma$ characterizes the diffusion rate of the random process.

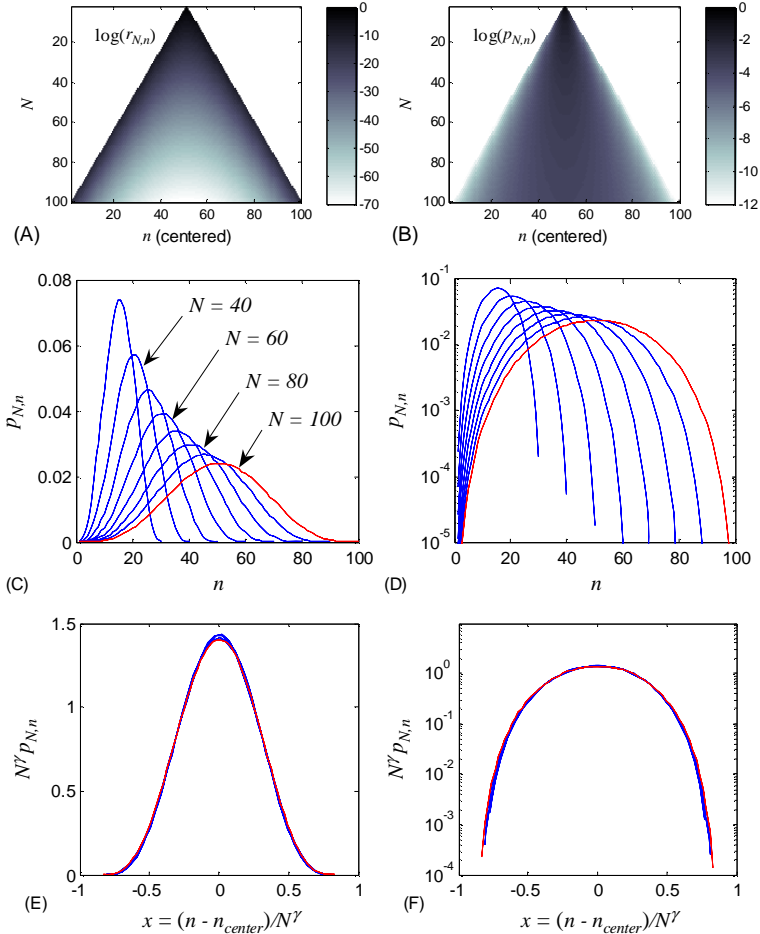

Fig. 4: The behavior of the MTG model with parameter $q_{corr} = 0.8$. Parts (A)-(B) show reduced probabilities $\log(r_{N,n})$ and probabilities $\log(p_{N,n})$ up to $N = 100$ as grayscale images. Parts (C)-(D) show uncentered and unscaled probabilities for $N = 30, 40, \ldots, 100$, in both linear and log units. Parts (E)-(F) show centered and scaled probabilities, illustrating the attractor in probability space. We find $\gamma = 0.88$.

*2.1 A q-describable model with $q_{ent} = 1$ and a q-Gaussian attractor*

We consider first the model referred to as MTG in Fig. 1, that has provided [5] numerical evidence for a generalized central limit theorem. The construction works in analogy to the case of independent subsystem composition, where $r_{N,0} = r_{1,0}^N \equiv p^N$. Application of the Leibnitz rule then yields the usual form $r_{N,n} = p^{N-n}(1-p)^n$. Global correlations can be introduced through the *q*-product [10], defined as follows

$$x \otimes_q y \equiv \left(x^{1-q} + y^{1-q} - 1\right)^{\frac{1}{1-q}} \qquad (x \otimes_1 y = xy), \tag{8}$$

where it is required that $x, y \geq 1$, and $q \leq 1$. From this definition, the form

$$\frac{1}{r_{N,n}} = \left(\frac{1}{p}\right)^N, \tag{9}$$

valid for independent subsystem composition, is generalized to

$$r_{N,0} = \left[\left(\frac{1}{p}\right) \otimes_{q_{corr}} \left(\frac{1}{p}\right) \otimes_{q_{corr}} \cdots \otimes_{q_{corr}} \left(\frac{1}{p}\right)\right]^{-1} = \left[Np^{q_{corr}-1} - (N-1)\right]^{\frac{1}{q_{corr}-1}} \tag{10}$$

where *corr* stands for *correlation*. The case $q_{corr} = 1$ yields the independent case, as the *q*-product reduces to the usual product when $q = 1$. And the case $q_{corr} = 0$ and $p = 0.5$ yields $r_{N,0} = (N+1)^{-1}$, which is the original Leibnitz triangle construction [11]. The MTG model is illustrated in Figs. 4-5 for $q_{corr} = 0.8$, and 0.3, respectively.

With the $r_{N,0}$ specified as in Eq. 10 above, the Leibnitz rule is applied to define the remaining reduced probabilities. As discussed in more detail in the original reference [5], this model reproduces, in the limit $N \to \infty$, a *q*-generalization of the de Moivre-Laplace version of the Central Limit Theorem. Indeed, the probability distribution closely approaches a *q*-Gaussian,

$$p(x) \propto e_Q^{-\beta x^2}, \tag{11}$$

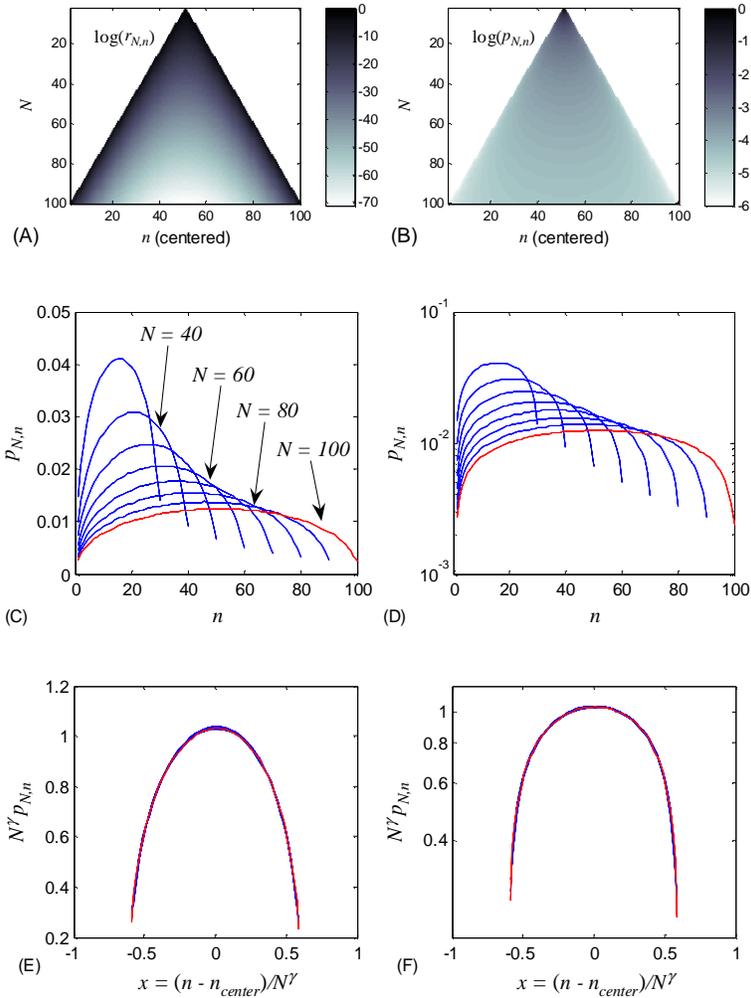

Fig. 5: The behavior of the MTG model with parameter $q_{corr} = 0.30$. Parts (A)-(B) show reduced probabilities $\log(r_{N,n})$ and probabilities $\log(p_{N,n})$ up to $N = 100$ as grayscale images. Parts (C)-(D) show uncentered and unscaled probabilities for $N = 30, 40, \ldots, 100$, in both linear and log units. Parts (E)-(F) show centered and scaled probabilities, illustrating the attractor in probability space. We find $\gamma = 0.96$.

except for a very small amount of asymmetry. The relation between the original $q_{corr}$ supplied to the construction (via the $q$-product of Eq. 10), and the value $Q$ that characterizes the limiting $q$-Gaussian distribution function is quite simple, and obtained numerically [5] as $Q = 2 - 1/q_{corr}$.

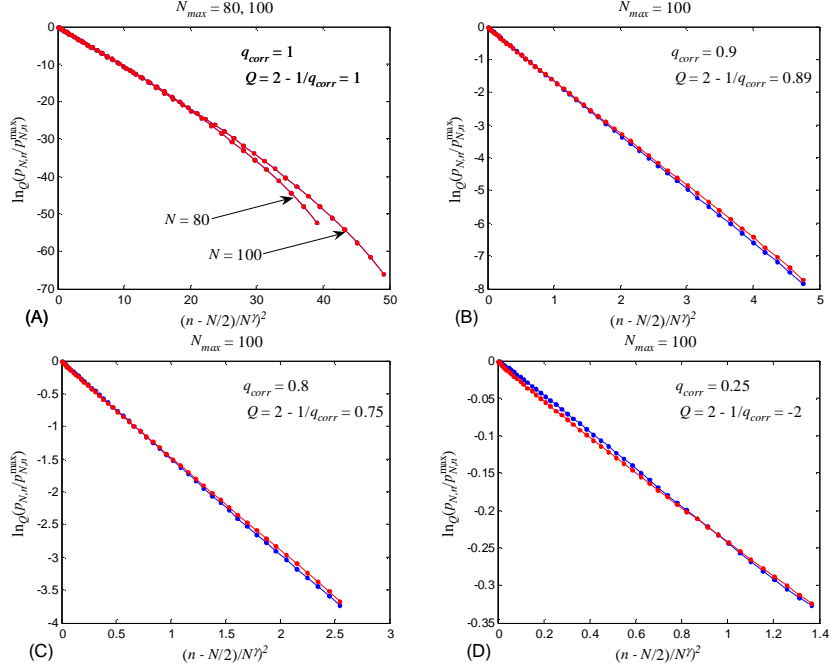

Fig. 6: The probabilities $p_{N,n}$ for the MTG model, replotted to illustrate the approach to the $q$-Gaussian form. (A) $q_{corr} = 1.0$, where the deviation from a straight line is a finite-size effect, also apparent in Fig. 3F, and is illustrated using two $N$ values, (B) $q_{corr} = 0.9$, (C) $q_{corr} = 0.8$, as shown in Fig. 4, (D) $q_{corr} = 0.25$, similar to the case shown in Fig. 5. In each case the $Q$-log of the probabilities, with the appropriate $Q$, is plotted as a function of scaled $n^2$, showing a straight line, in agreement with the $q$-Gaussian form.

Figure 6 illustrates the $q$-Gaussian attractor explicitly by plotting the $q$-log of the probability set vs. the square of the centered index $|n - n_{center}|$, which yields a straight line for $q$-Gaussian distributions. Indeed, for the independent case of Fig. 3, where $q_{corr} = 1$, and the various $q_{corr}$ values of Fig. 6, we find good numerical agreement with the equation $Q = 2 - 1/q_{corr}$. The slight asymmetry noted in Ref. [5] is also apparent in Fig. 6B-D.

The correlations introduced into this model are, although global in nature, relatively weak, and the model exhibits $q_{ent} = 1$ for all values of $q_{corr}$. This may be expected, since $W^{eff} \sim 2^N$. Figure 7 shows an example of $S_q$ vs. $N$ for $q_{corr} = 0.25$, demonstrating that $q_{ent} = 1$.

### 2.2 *A q-describable model with $q_{ent} = 1$, and non-q-Gaussian attractor*

We consider next a generalization of the independent system case, referred to as TGS1 in Fig. 1, which has appeared previously in Ref. [4]. This model is defined using a stretched exponential as follows

$$r_{N,0} = r_{1,0}^{N^\alpha} = p^{N^\alpha}, \tag{12}$$

where $\alpha$ is a parameter constrained to the range [0, 1] (for $\alpha$ outside this range, some of the $r_{N,n}$ become unphysical). The other reduced probabilities are constructed using the Leibnitz rule. At one extreme, $\alpha = 0$, only two values, $r_{N,0}$ and $r_{N,N}$ are nonzero. The opposite extreme, $\alpha = 1$ reduces to the independent case (Gaussian attractor). Figure 8 illustrates the behavior of this model for case $\alpha = 0.9$.

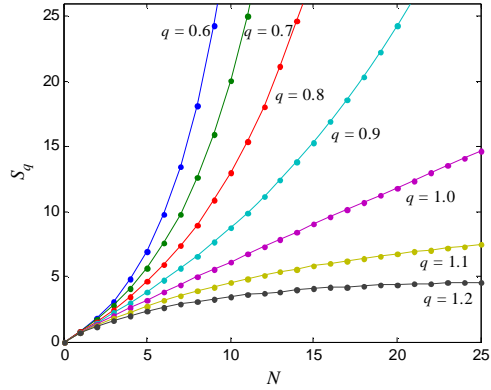

Fig. 7: Entropy growth in the model "MTG" defined in Eq. 10, for the case $q_{corr} = 0.25$, showing that $S_{BGS}$ is extensive (i.e., $q_{ent} = 1$).

The attractor clearly deviates from the Gaussian, however for the entire range of $\alpha$ this model exhibits extensive $q_{ent} = 1$. In Fig. 1B, this model thus resides in the $q_{ent} = 1$ region of the $q$-describable systems, and terminates at the Gaussian attractor, thus providing a second example of a system for which the attractor deviates from the Gaussian, yet the correlations are not strong enough to give $q_{ent} \neq 1$.

### 2.3  *Two q-describable models with $q_{ent} \neq 1$, and non-q-Gaussian attractor*

We consider next a $q$-describable model [4] for which $q_{ent} \neq 1$. This model is referred to as model TGS2 in Fig. 1B, and is shown as a series of dots, since the model is discretely parameterized. The model is based on restricted occupancy, such that the effective occupancy $W^{eff} \sim N^d$ for some positive integer $d$. The reduced probabilities are specified as

$$r_{N,n} = \begin{cases} \dfrac{1}{W^{eff}} = \left( \displaystyle\sum_{k=0}^{\min(d,N)} \dfrac{N!}{(N-k)!k!} \right)^{-1} & n \leq d \\ 0 & \text{otherwise} \end{cases}. \qquad (13)$$

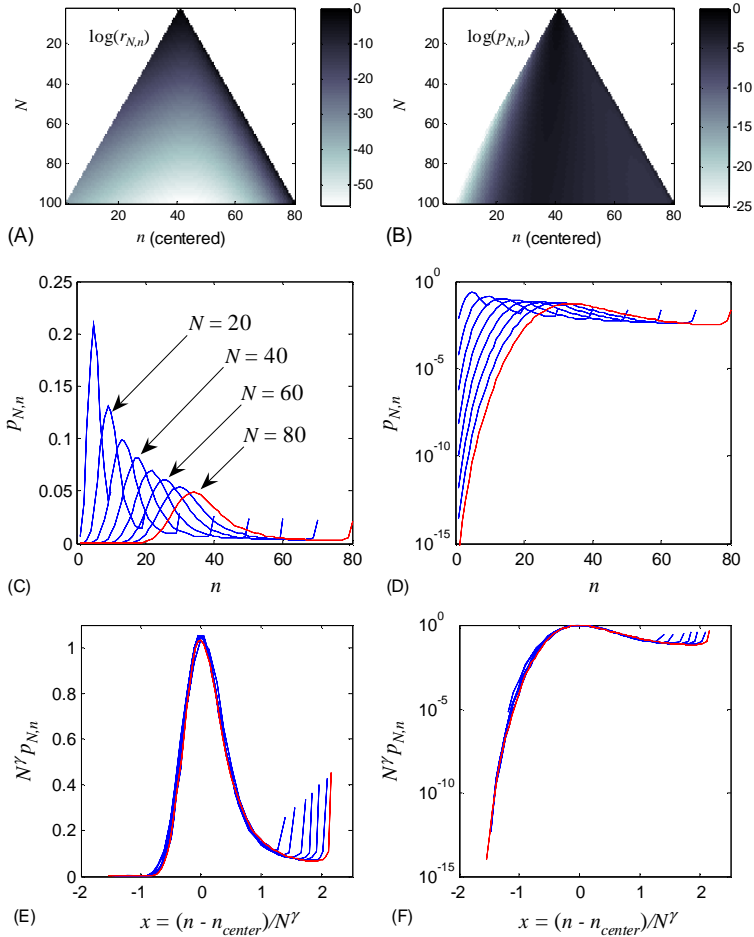

Fig. 8: Behavior of the model "TGS1", defined in Eq. 12, for $\alpha = 0.9$. Parts (A)-(B) show reduced probabilities $\log(r_{N,n})$ and probabilities $\log(p_{N,n})$ up to $N = 100$ as grayscale images. Parts (C)-(D) show uncentered and unscaled probabilities for $N = 10, 20, \ldots, 80$, in both linear and log units. Parts (E)-(F) show centered and scaled probabilities, illustrating the attractor in probability space. We find $\gamma = 0.70$.

In case $N \leq d$, the summation extends to $k = N$, so the probabilities reproduce the case of independent system composition. For $N > d$, the summation extends only to $k = d$, yielding non-zero occupation only in the first $d+1$ reduced probabilities. Note the normalization condition of Eq. 4 is satisfied by this prescription. Figure 9 illustrates the behavior of this model, for the case $d = 15$.

In the case $d = N$, this model reduces to the independent case, yielding a Gaussian attractor, and $q_{ent} = 1$. For finite $d$, the attractor of this model is a delta function, clearly deviating from the Gaussian distribution. Furthermore, because [4] the effective number of states $W^{eff} \sim N^d$, the value of $q_{ent}$ satisfies the theoretical form $q_{ent} = 1 - 1/d$. Figure 10 illustrates this agreement with numerical simulation results, in which the $q_{ent}$ was determined in each case by fitting a straight line to the $S_q$ vs. $N$ curves, and maximizing the $R^2$ value of the fit. Figure 11 gives an example of the entropy growth curves for the case $d = 5$, verifying that $q_{ent} = 0.8$ gives linear entropy growth, as predicted.

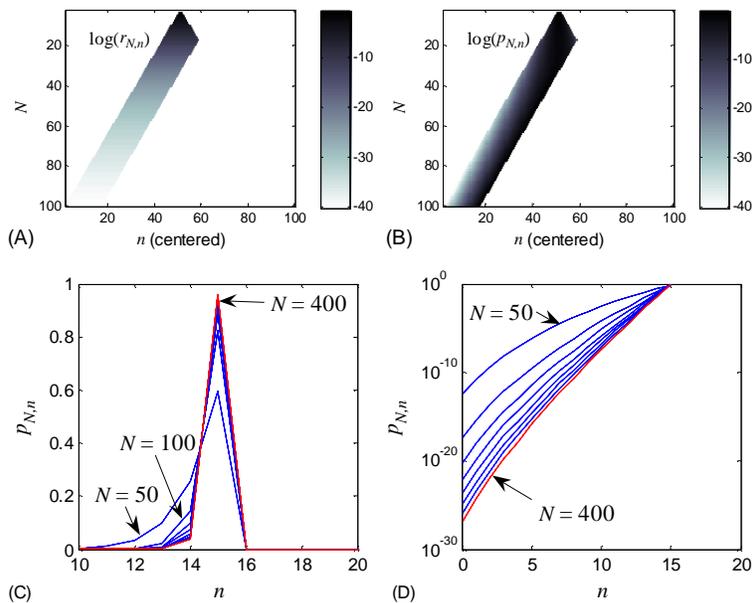

Fig. 9: The behavior of the TGS2 model with parameter $d = 15$. Parts (A)-(B) show reduced probabilities $\log(r_{N,n})$ and probabilities $\log(p_{N,n})$ up to $N = 100$ as grayscale images. Parts (C)-(D) show uncentered and unscaled probabilities for $N = 50, 100, \ldots, 400$, in both linear and log units. We find $\gamma = 0$.

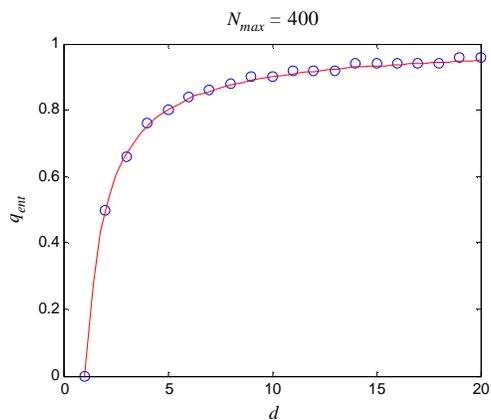

Fig. 10: The value of $q_{ent}$ as a function of $d$ for the model "TGS2" defined in Eq. 13. Numerical data were obtained by analyzing the linearity of $S_q(N)$ vs. $N$ plots, for $N$ ranging up to $N_{max} = 400$. The open circles represent numerical data, and the solid curve is the theoretical result $q_{ent} = 1 - 1/d$.

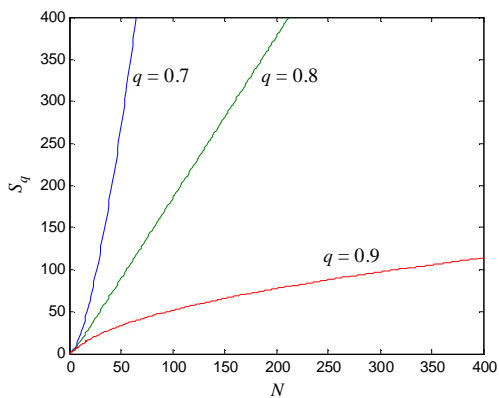

Fig. 11: Entropy growth for the model TGS2 defined in Eq. 13, for the case $d = 5$, demonstrating numerically that linear entropy growth is observed for $q_{corr} = 1 - 1/d = 0.8$.

The model TGS2 does not show scale invariance, as the Leibnitz rule is violated, even asymptotically. However, an asymptotically scale-free variation on this model, using the same cutoff parameter $d$, has been produced in Ref. [4]. This model is referred to as TGS3 in Fig. 1B, and has similar properties to that of TGS2, being discretely parameterized, exhibiting a similar delta function attractor, and satisfying the same theoretical prediction $q_{ent} = 1 - 1/d$. Figure 12 illustrates the behavior of this model for the case $d = 4$.

Both TGS2 and TGS3 provide examples of systems for which the global correlations are so strong that $q_{ent} \neq 1$, in addition to the attractor deviating substantially from the Gaussian.

### 2.4 *A new class of q-describable models with $q_{ent} \neq 1$, and non-q-Gaussian attractor*

We introduce now a new class of models which offer further insight into the conditions necessary for $q_{ent} \neq 1$. Three examples from this class will be presented.

The first example from this new class, referred to as "present1" in Fig. 1B, is defined by the reduced probabilities

$$r_{N,n} = \frac{N^{\alpha n}}{\sum_{k=0}^{N} \frac{N!}{(N-k)!k!} N^{\alpha k}} \tag{14}$$

where $\alpha$ is a real constant, and the denominator ensures proper normalization. This model introduces correlations so strongly that the probability set at level $N$ does not depend in a simple way on the probability set at level $N-1$, and so does not adhere to the Leibnitz rule. The case $\alpha = 0$ reduces to the independent composition case, and so yields $q_{ent} = 1$. However, for $\alpha \neq 1$, we find $q_{ent} \neq 1$. Indeed, Fig. 13 shows the entropy growth with $N$ for $\alpha = -1.0$, and for several values of $q$, demonstrating $q_{ent} \approx 0.7$. Figure 14 shows the dependence of $q_{ent}$ on the parameter $\alpha$, where each value of $q_{ent}$ is determined by maximizing the $R^2$ value from straight line fits to $S_q(N)$ vs $N$. The model is symmetric in $\alpha$, as shown. The figure also shows a fit to a stretched $q$-Gaussian of the form

$$q_{ent} = e_Q^{-\alpha^4} = \left(1 - (1-Q)\alpha^4\right)^{\frac{1}{1-Q}}, \tag{15}$$

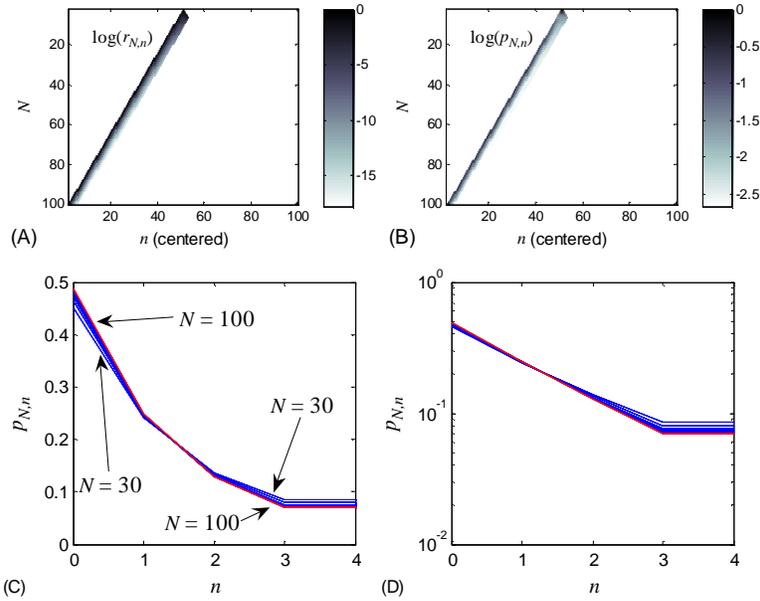

Figure 12: The behavior of the "TGS3" model with parameter $d = 4$. Parts (A)-(B) show reduced probabilities $\log(r_{N,n})$ and probabilities $\log(p_{N,n})$ up to $N = 100$ as grayscale images. Parts (C)-(D) show uncentered and unscaled probabilities for $N = 30, 40, \ldots , 100$, in both linear and log units. We find $\gamma = 0$.

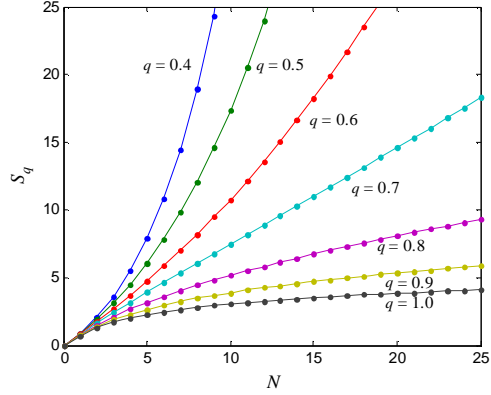

Fig. 13: Entropy growth for the model "present1" defined in Eq. 14, with parameter $\alpha = -1$. The entropy growth curves for various $q$, showing the entropy $S_q$ is extensive for $q_{ent} \approx 0.7$.

where the fitted parameter $Q \cong 5.0$. This parameter is found to depend in a systematic way on the maximum value $N$ used in the straight-line fits to $S_q(N)$ vs. $N$, and is found to closely approach $Q = 5.0$ as $N \to \infty$. The asymptotic behavior for large $\alpha$ is thus well-described by the simple form $q_{ent} \sim |\alpha|^{-1}$.

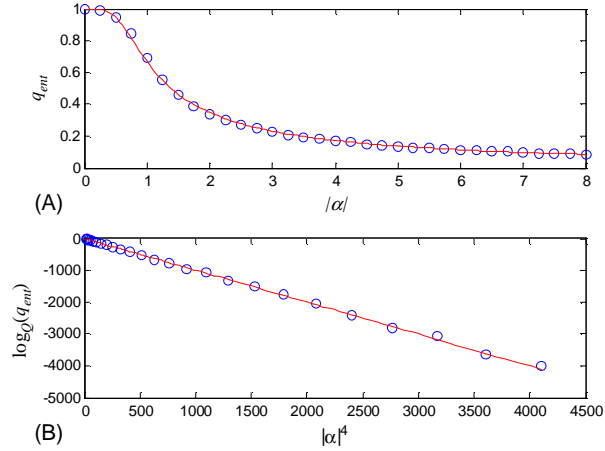

Fig. 14: Extensive $q_{ent}$ plotted as a function of the parameter $\alpha$, for the model "present1" defined in Eq. 14. (A) Linear axes, and (B) scaled axes. The open circles represent data obtained numerically, and the solid line is a fit to a stretched $q$-Gaussian, characterized by $Q = 4.95 \approx 5$. As expected, we find $q_{ent} = 1$ when $\alpha = 0$. The extensive $q_{ent}$ was determined by analyzing the linearity of $S_q(N)$ plot for $N$ ranging up to $N_{max} = 25$.

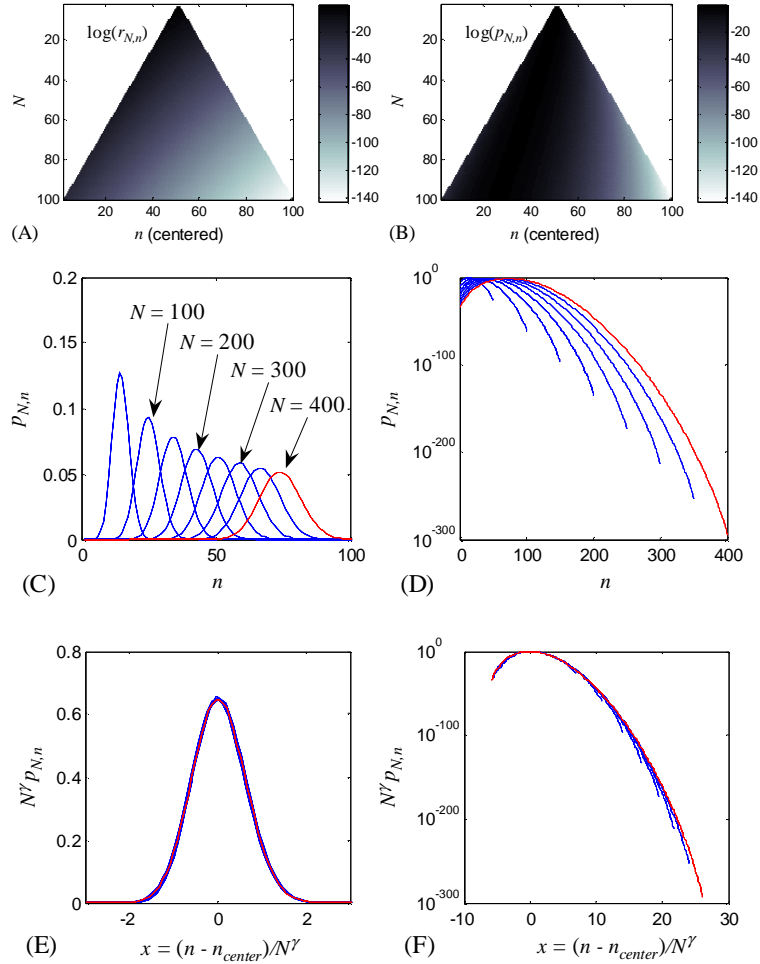

Fig. 15: The behavior of the "present1", defined in Eq. 14, with $\alpha = -0.25$. Parts (A)-(B) show reduced probabilities $\log(r_{N,n})$ and probabilities $\log(p_{N,n})$ up to $N = 400$ as grayscale images. Parts (C)-(D) show uncentered and unscaled probabilities for $N = 50, 100, \ldots, 400$, in both linear and log units. Parts (E)-(F) show centered and scaled probabilities, illustrating the attractor in probability space. We find $\gamma = 0.42$.

This model provides an example of a system for which $q_{ent} \neq 1$ and $W^{eff}(N)$ is not explicitly constructed to grow as a power of $N$ (as in TGS2 and TGS3, discussed above). The behavior of this model can be understood qualitatively as follows. Consider the case $\alpha < 0$, in which case each reduced probability $r_{N,k}$ dominates the next one $r_{N,k+1}$ by a factor of $N^\alpha$. As $N$ grows, the first two probabilities $p_{N,0}$ and $p_{N,1}$ are increasingly dominant, becoming approximately $1 - \varepsilon$ and $\varepsilon$, with $\varepsilon \sim N^{1+\alpha}$. Clearly, as $N$ increases, the BGS entropy of this probability set *decreases* towards zero. A small value $q_{ent}$ tends to emphasize the very small probabilities $p_{N,n}$ for $n \geq 2$, effectively drawing the calculation away from the zero entropy point, and ensuring the entropy $S_q$ increases linearly.

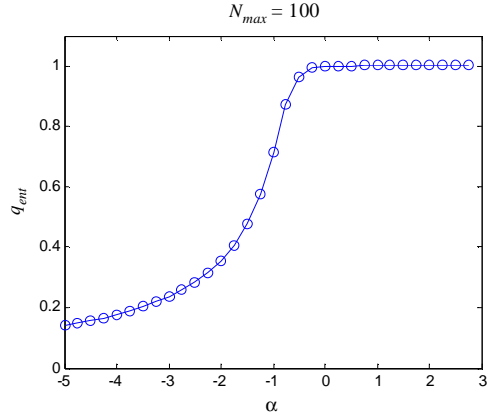

Fig. 16: Numerical results for extensive $q_{ent}$ plotted as a function of the parameter $\alpha$, for the model "present2" defined in Eq. 16. The symmetry of the model "present1" in $\alpha$ has been broken, and $q_{ent} = 1$ for all $\alpha \geq 0$. The solid line is a guide for the eye. The values $q_{ent}$ for $\alpha < 0$ coincide with those of Fig. 14.

The attractor for this model for $\alpha = -0.25$ is shown in Fig. 15. As $\alpha$ is varied away from 0 (the independent case), we find the diffusion exponent varies systematically away from the normal case $\gamma = 0.5$, and $\gamma \to 0$ as $\alpha \to 0$, as can be seen in Fig. 20. For $|\alpha| > 1$, the exponent $\gamma$ remains 0, indicating a localized state, and consistent with the picture of only two probabilities ($P_{N,0}$ and $p_{N,1}$) which are appreciably non-zero.

We now introduce a symmetrized version of this model, referred to as "present2" in Fig. 1B, and defined by

$$r_{N,n} = \frac{N^{\alpha n'}}{Z} \qquad (16)$$

where

$$n' = \begin{cases} n & n \leq \lfloor N/2 \rfloor \\ N - n & otherwise \end{cases} \quad n = 0, \ldots, N \qquad (17)$$

and where $Z$ is a constant chosen to satisfy normalization. With this definition, $\alpha > 0$ ($\alpha < 0$) corresponds to reduced probabilities that reach a minimum (maximum) at $n = N/2$. When $\alpha \geq 0$, this model exhibits $q_{ent} = 1$. When $\alpha < 0$,

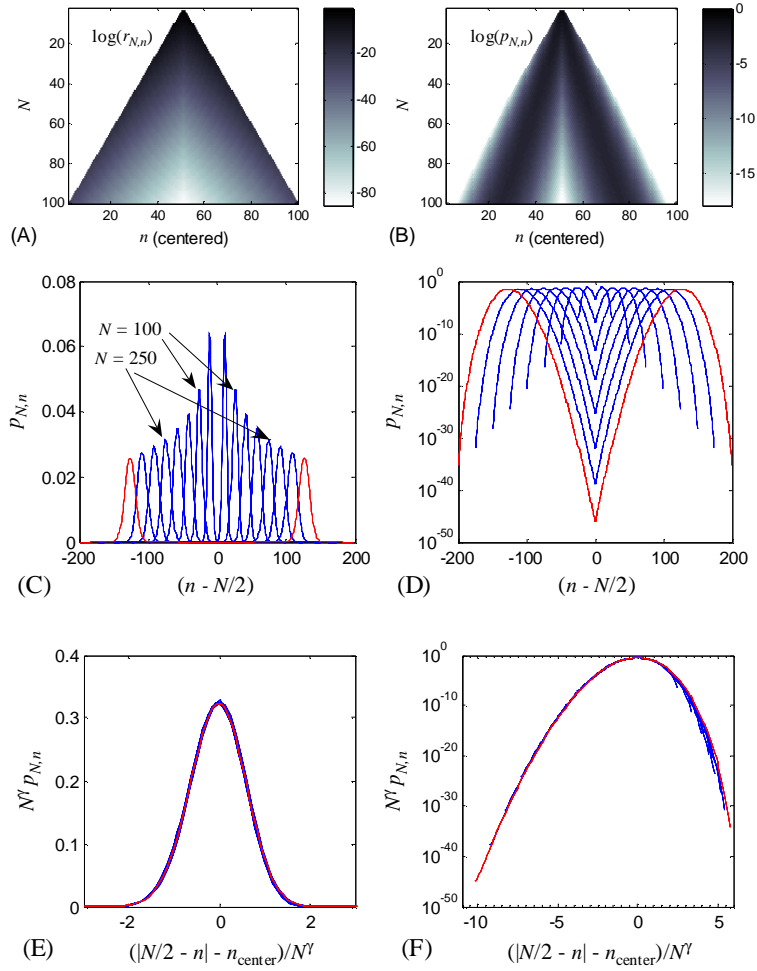

Fig. 17: The behavior of the "present2", defined in Eq. 16, with $\alpha = -0.25$. Parts (A)-(B) show reduced probabilities $\log(r_{N,n})$ and probabilities $\log(p_{N,n})$ up to $N = 400$ as grayscale images. Parts (C)-(D) show uncentered and unscaled probabilities for $N = 50, 100, \ldots, 400$, in both linear and log units. Parts (E)-(F) show centered and scaled probabilities, illustrating the attractor in probability space. We find $\gamma = 0.42$.

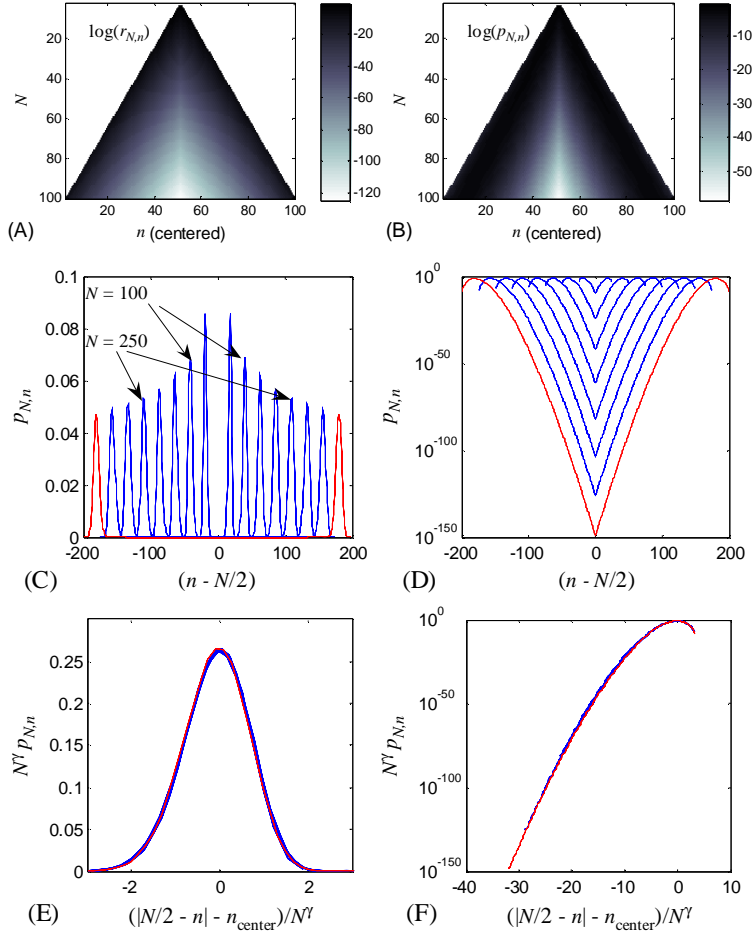

Fig. 18: The behavior of the "present2", defined in Eq. 16, with $\alpha = -0.5$. Parts (A)-(B) show reduced probabilities $\log(r_{N,n})$ and probabilities $\log(p_{N,n})$ up to $N = 400$ as grayscale images. Parts (C)-(D) show uncentered and unscaled probabilities for $N = 50, 100, \ldots, 400$, in both linear and log units. Parts (E)-(F) show centered and scaled probabilities, illustrating the attractor in probability space. We find $\gamma = 0.28$.

$q_{ent}$ behaves quantitatively like the model "present1" defined in Eq. 14 above. However, as $\alpha$ decreases we expect to see a difference, because the smallest $p_{N,n}$ in this model ("present2") are not as small as the smallest $p_{N,n}$ in the model "present1". The behavior of $q_{ent}$ as a function $\alpha$ is shown in Fig. 16. Due to the symmetry of this model about $N/2$, the attractor for this model has two branches, as shown for $\alpha = -0.25$ and $\alpha = -0.5$ in Figs 17 and 18, respectively, and is scaled with ordinate measured from the point of symmetry.

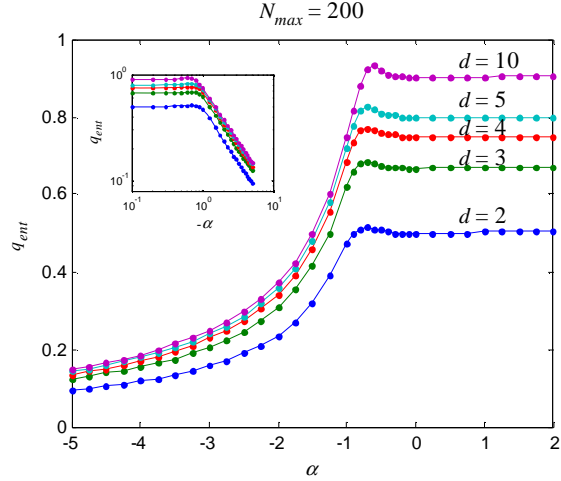

Fig. 19: Numerical results for extensive $q_{ent}$ plotted as a function of the parameter $\alpha$, for the model "present3", defined in Eq. 18. Shown are plots for $d = 2, 3, 4, 5,$ and $10$. For $\alpha \geq 0$, we observe $q_{ent} = 1 - 1/d$, as predicted based on reduced occupancy. For $\alpha < 0$, rapidly decreasing probabilities, coupled with explicitly reduced occupancy, leads to behavior much like that in the model "present 1", where $q \sim |\alpha|^{-1}$ as $\alpha \to -\infty$. In the region $-1 < \alpha < 0$, we find the two effects in competition, leading to more complex behavior. The inset contains a log-log plot of the negative $\alpha$ portion of the data, showing explicitly that $q_{ent} \sim |\alpha|^{-1}$, and the sharp change in behavior at $\alpha = -1$.

Finally, we introduce a version of this model, referred to as "present3" in Fig. 1B, with a cutoff, similar to the cutoff introduced in the model TGS2. The model is defined by

$$r_{N,n} = \begin{cases} \dfrac{N^{\alpha n}}{Z} & n \leq d \\ 0 & \text{otherwise} \end{cases} \quad (18)$$

where $Z$ is a quantity (depending on $N$) chosen to ensure proper normalization. When $\alpha \geq 0$, this model behaves like the model TGS1, with $q_{ent} = 1 - 1/d$. This similarity holds even though the occupation probabilities are markedly different from those of TGS1. We thus find the behavior is dominated by $W^{eff}$, which scales as $W^{eff} \sim N^d$. When $\alpha < 0$, the effective number of states is further restricted due to the sharply decreasing values of $r_{N,n}$, and $q_{ent}$ behaves much like the models "present1" and "present2" above, which, as $\alpha \to -\infty$, require $q_{ent} \to 0$. These results are shown in Fig. 19.

## 3. Conclusion

A number of simple models have been presented that illustrate the influence of global correlations on the attractor in probability space, and on the extensivity of the generalized entropy $S_q$. In each model studied, even weak global correlations (parametrically induced) affect the attractor, leading to deviations from the Gaussian form associated to independent systems, and representative of the usual Central Limit Theorem. Weak global correlations do not, however, necessarily cause the extensivity of the system to appear anomalous. Indeed, the models MTG, TGS1, and "present2" with $\alpha > 0$ have $q_{ent} = 1$, whereas the models TGS2, TGS3, "present1", "present2" with $\alpha < 0$, and "present3" have $q_{ent} \neq 1$.

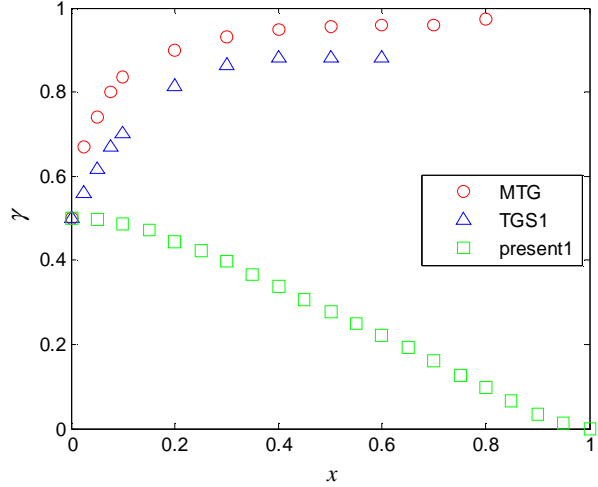

Fig. 20: Diffusion exponent $\gamma$ vs. model parameters $x$, for the models MTG, TGS1, and "present1". For all three models, $x = 0$ corresponds to the independent case, with normal diffusion exponent $\gamma = 0.5$. In the model MTG $x = 1 - q_{corr}$, and in TGS1 $x = 1 - \alpha$, each of these two models exhibiting *superdiffusion*. In the model "present1," $x = |\alpha|$, and the model exhibits *subdiffusion*. The values $\gamma = 0, 1$ respectively correspond to localization and ballistic superdiffusion.

Another connection worth noticing concerns the diffusion exponent $\gamma$ which characterizes how "space" scales with $N$. In the absence of global correlations we have $\gamma = \frac{1}{2}$ (i.e., normal diffusion). In the presence of global correlations we have $\gamma \neq \frac{1}{2}$. The case $0 \leq \gamma < 1/2$ (i.e., subdiffusion) emerges with both $q_{ent} = 1$ (model "present2" with $\alpha > 0$) and $q_{ent} \neq 1$ (models TGS2, TGS3, "present1", "present2" with $\alpha < 0$, and "present3"). The case $\gamma > \frac{1}{2}$ (i.e., superdiffusion) has only emerged for $q_{ent} = 1$ (models MTG, and TGS1). None of the present models has simultaneously exhibited $\gamma > \frac{1}{2}$ and $q_{ent} \neq 1$, but we see no reason for such a situation to be excluded a priori. For the particular case of subdiffusion, we have obtained $\gamma = 0$ (i.e., localization) every time that the probabilistic model had, for increasingly large $N$, probabilities sensibly different from zero only for a *finite* width in $n$ (hence for the models TGS2, TGS3, "present2" with $\alpha > 0$, and "present3"). Finally, all the cases for which $\gamma$ continuously varies with the model parameter characterizing global correlation are indicated in Fig. 20.

The present results can be considered to be a first step in the understanding of the interconnections between (i) the entropic extensivity, (ii) the attractor in the sense of a central limit theorem, (iii) the type of diffusion, and (iv) the distribution which, under appropriate constraints, extremizes the entropic functional. In the *absence of global correlations* (and assuming that the variances of the distributions that are being composed are *finite*, which always is the case for binary random variables) we have respectively $q_{ent} = 1$, a Gaussian attractor, normal diffusion, and the maximization of $S_{BGS}$ yields a Gaussian. The challenge at this stage of course is to attain a similar understanding in the *presence of global correlations*, either (strictly or asymptotically) scale-invariant or not.

### Acknowledgements

We have benefited from interesting discussions with S. Umarov, R. Hersh, S. Steinberg, K. Nelson, W. Thistleton, A. Williams and M. Gell-Mann. JM and CT gratefully acknowledge support from A. Williams of the Air Force Research Laboratory, Information Directorate, and SI International, under contract No. FA8756-04-C-0258. LM acknowledges support from PRONEX and CNPq (Brazilian Agencies).